\documentclass{article}

\usepackage[margin=.75in]{geometry}

\usepackage{graphicx}
\usepackage{cite}

\begin{document}
	
\newcommand{\EQ}{Eq.~}
\newcommand{\EQS}{Eqs.~}
\newcommand{\FIG}{Fig.~}
\newcommand{\FIGS}{Figs.~}
\newcommand{\SEC}{Sec.~}
\newcommand{\SECS}{Secs.~}

\begin{center}
{\Large Spatial prisoner's dilemma optimally played in
small-world networks}
\bigskip

${}^1$ Naoki Masuda and ${}^2$ Kazuyuki Aihara\\
${}^1$ Department of Complexity Science and Engineering,\\
Graduate School of Frontier Sciences, the
University of Tokyo,\\
7--3--1, Hongo, Bunkyo-ku, Tokyo, 113--8656, Japan\\
${}^2$
Core Research for Evolutional Science and Technology, Japan Science
and Technology Corporation,\\
4--1--8, Hon-cho, Kawaguchi, Saitama, 332--0012, Japan\\
%
\end{center}
\bigskip

\begin{abstract}
Cooperation is commonly found in ecological and social systems even when it
apparently seems that individuals can benefit from selfish behavior.
We investigate how cooperation emerges with the spatial prisoner's
dilemma played in a class of networks ranging from regular lattices to
random networks. We find that, among these networks, 
small-world topology is the
optimal structure when we take into
account the speed at which cooperative behavior propagates.
Our results may explain why the small-world
properties are self-organized in real networks.
\end{abstract}

\bigskip

 

\section{Introduction}\label{sec:introduction}

Cooperation among individuals is everywhere in ecological and social
systems such
as animal groups and human societies.  How altruistic behavior emerges in the
situations where each individual is apparently tempted to defect 
has been debated for a long time. In the game theory,
this situation is typically formulated as the prisoner's dilemma (PD)
\cite{Axelrod81,Axelrod84}. To resolve the dilemma and explain the actually
found altruism, various mechanisms such as kin selection
\cite{Hamilton}, reciprocal altruism in iterated games
\cite{Axelrod81,Axelrod84}, and spatial games
\cite{Axelrod84,Nowak92,Nowak93ijbc,Nowak94pnas,Nowak94ijbc} have been
proposed.

For example, the iterated PD with evolutionary dynamics assumes a population
of players randomly interacting with each other. A round of interaction
between two players consists of the repetition of PD, in which
a next PD is played with probability $w$ where $0<w<1$.
Consequently, they are provided with the payoff summed over the round.
After playing one round of the game with all
the other players, the
players that have gained higher total payoff are more likely to survive in the next
round.  When $w$ is sufficiently large, cooperative strategies such as
Tit-for-Tat outperform the mean players such as pure defectors.
Only the former can elicit reciprocal altruism from other good players
to benefit both of them and is not exploited by mean players
\cite{Axelrod81,Axelrod84}.

Another solution is to introduce spatial structure as found in the
real world
so that a player interacts with a limited number of neighbors
in each round. If defection is not extremely advantageous,
clusters of cooperators can survive in the sea of defectors when
the network is clustered, that is, when one can expect that a friend
of my friend is another friend of mine with a significantly 
high probability.
Clustered cooperators of a certain size reciprocate each other to
surpass defectors that get payoff just by exploiting cooperators.
Two-dimensional square lattice networks each of whose
vertex has immediate eight neighbors (Moore neighborhood),
for example, are equipped with the clustering property and facilitate altruism
\cite{Hauert,Nowak92,Nowak93ijbc,Nowak94pnas,Nowak94ijbc}. 
Simple good strategies such as pure cooperators 
can be territorially stable in the noniterated PD even if they cannot
persist in random networks \cite{Axelrod84}. The results have also
been extended to more general cases involving stochasticity,
asynchronous updating, irregularity in lattices
\cite{Nowak94pnas,Nowak94ijbc}, and the spatial Hawk-Dove games
\cite{Killingback96}. Moreover, interesting phenomena such as chaotic
fluctuation in the proportion of cooperators and dynamic fractals of
population patterns are reported with this framework
\cite{Nowak92,Nowak93ijbc}.

Generally speaking, 
the network topology significantly influences the dynamical behavior
in ecological and social networks
\cite{Abramson,Albert,Kim,Kuperman,Wattsbook}.
Regular lattices employed in spatial games have an important
feature of social networks, that is, the cluster property that enables
a group of players to cooperate and benefit each other \cite{Hauert}.
Randomly connected networks lack this property. On the other hand, real
networks also have the property that, on the average, any two
individuals are connected via a path much shorter than is expected for
regular lattices \cite{Albert,Newman01,Watts,Wattsbook}. In a
relationship network among individuals,
one mainly interacts with other individuals in its own cluster or group, but
it also tends to interact with some individuals that belong to other groups.
The small-world properties comprising the clustered property and the
average short path length are realized in the pioneering mathematical
model by Watts and Strogatz, which suitably describes many
types of biological, physical, and social networks
\cite{Watts,Wattsbook}. To generate one-dimensional small-world
networks, let us start with a ring of $n$ vertices each of which has
$k/2$ nearest neighbors on each side. Then we rewire a proportion $p$
of the total edges by removing $pkn/2$ edges and creating $pkn/2$ new
edges each of whose initial vertex is the initial vertex of a removed
edge and its terminal vertex is randomly chosen so that the generated
graph does not have multiple edges or once removed edges.
Graphs with small positive rewiring probabilities
satisfy the small-world properties with small
average path lengths $L(p)$ and large clustering coefficients $C(p)$,
where $L(p)$ is defined as the shortest path
length between two points averaged over all point pairs, and
$C(p)$ is the normalized number of triangles 
\cite{Watts,Wattsbook}. The
regular lattice ($p=0$) and the random graph ($p=1$) are generated by
this procedure as the two extremes. 

The spatial PD in small-world networks, which may be more plausible
models for social interaction, has been examined by a few
authors. Watts \cite{Wattsbook} showed that generalized Tit-for-Tat
and Win-Stay-Lose-Shift strategies survive in the population mixed
with defectors if $C(p)$ is large enough, in other words, if $p$ is
small enough.  There exists a critical rewiring probability $p_c$
above which good strategies die out. As clusters collapse with an
increasing $p$, similar kinds of transition appear in epidemic models
with small-world networks \cite{Kuperman}.  Watts has also measured
the time necessary for the convergence to stable proportions of good
strategies, and related it to $L(p)$.  In \cite{Kim}, an influential
node is endowed with an increased number of shortcuts, and how
instantaneous strategy change of the influential node affects the
macroscopic dynamics is studied. In another work, proportions of
defectors in the small-world spatial PD is examined
\cite{Abramson}. In this Letter, we examine
the spatial PD in small-world networks and find
how the combined effects of $C(p)$ and $L(p)$
result in emergence and development of cooperation.  Especially, we
show that the proportion of cooperators determined by $C(p)$ and the
speed of convergence determined by $L(p)$ can be balanced, but
rapid convergence and total dominance of cooperators, though it
may seem socially preferable, are not simultaneously realized even in
the small-world regime. Instead, small-world networks realize
rapid emergence of slightly
suboptimal states with many cooperators in a global scale.

\section{Models and Results}

We assume that each vertex of a network, which ranges from a regular
lattice to a random graph by changing $p$, is occupied by a player.
In each round, a player interacts with its immediate neighbors. We
consider only memoryless strategies; each player chooses just
cooperation (denoted by $C$) or defection (denoted by $D$) in each
round \cite{Nowak92,Nowak93ijbc,Nowak94pnas,Nowak94ijbc}. When a
player chooses $C$, it receives payoff $R$ (reward) or $S$ (sucker) as
the opponent chooses $C$ or $D$, respectively.  A player that chooses
$D$ receives $T$ (temptation) or $P$ (punishment) as the opponent
chooses $C$ or $D$, respectively. Given $T>R>P>S$, a player is always
tempted to defect no matter whether the opponent takes $C$ or $D$. The
combination of $D$ and $D$, with which both of the players get
unsatisfactory payoff $P$, is the unique Nash equilibrium in a single
game.  Each player sums the payoff received by playing a single PD
with its neighbors and compares the sum with those of the
neighbors. Among them, the strategy with the maximal payoff is copied
as the player's strategy in the next round. This imitation procedure
is considered to stem from genetic evolution or social
learning. Following the earlier work, we assume $T=b>1$, $R=1$, and
$P=S=0$ \cite{Nowak92,Nowak93ijbc,Nowak94pnas,Nowak94ijbc}.  Though
the condition $P>S$ is violated, its influence on the dynamics is
negligible.

Let us denote the dimension of the underlying network by $d$, and 
we set $n=3600$ and $k=8$.
We examine the case of $d=1$ in which
$n$ vertices are aligned in a ring \cite{Abramson,Wattsbook,Pollock}
and the case of $d=2$ in which
60 $\times$ 60 vertices are aligned in a square lattice with
periodic boundary conditions. The connection structure is
parameterized by $p$.  When $p=0$, each vertex is connected to
four nearest neighbors on each
side for $d=1$ or to the eight players in the 
Moore neighborhood for $d=2$.  For
positive values of $p$, we randomly rewire the proportion $p$ of
the edges.  If the conventional rewiring procedure were used, the vertex
degree would vary from vertex to vertex after rewiring. To avoid
artificial normalization of payoff due to the dispersed number of
neighbors, we randomly rewire the edges keeping the degree of every
vertex constant \cite{Masuda_BC}.

\begin{figure}
\begin{center}
\includegraphics[height=2in,width=2in]{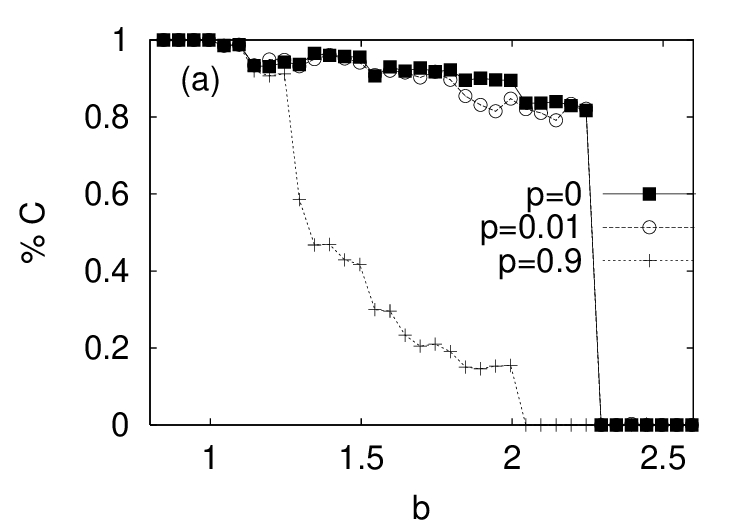}
\includegraphics[height=2in,width=2in]{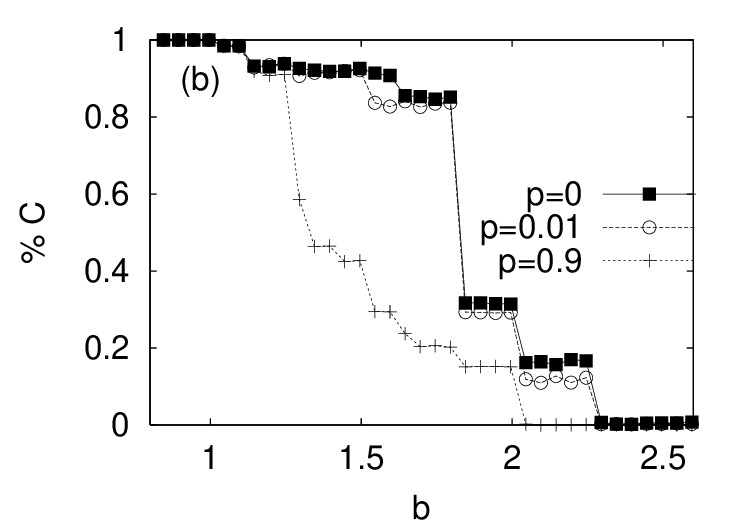}
\caption{The proportion of cooperators after transient for (a) $d=1$
and (b) $d=2$. We set $c(0) = 0.98$, and the statistics are based on averaging
over 10 trials.}
\label{fig:ratio_p_b}
\end{center}
\end{figure}

For various values of $p$ and $b$, the proportion of cooperators after
the transient is shown in \FIG\ref{fig:ratio_p_b}. The initial
proportion of cooperators that are randomly and independently chosen
for all the sites is equal to $c(0)=0.98$. It is known for $d=2$ and
$p=0$ that cooperators dominate for small $b$, defectors occupy the
whole space for sufficiently large $b$, and intermediate values of $b$
($1.8<b<2.0$) result in chaotic fluctuation \cite{Nowak92,Nowak93ijbc}.
Figure~\ref{fig:ratio_p_b}(b) shows that these regimes persist for any
values of $p$ although a larger value of $p$ lowers the critical
values of $b$ for the transitions.  Moreover, as we will see shortly,
we can identify three ranges of $b$ based on the combination of final
proportions of cooperators and dependence of the results on $p$.
These observations hold true also for $d=1$ as shown in
\FIG\ref{fig:ratio_p_b}(a) just with slight shifts of the critical
values of $b$ dividing different regimes.  Such kinds of quantitative
change in different dimensions are also observed
with regular lattices
\cite{Nowak94pnas,Nowak94ijbc}. The three regimes are as follows.

\begin{description}
\item[(i)]
For small $b$, it
is not so tempting for players to exploit cooperators. Consequently,
the proportion of cooperators converges to a value close to 1 
regardless of $p$.

\item[(ii)]
The number of cooperators highly depends on $p$ roughly for
$1.3\le b\le 2.3$.  In
this case, $C(p)$ needs to be larger for cooperators to survive
\cite{Wattsbook}.  The chaotic situation is also included in this
regime. 

\item[(iii)]
For larger $b$, players are inclined to betray.
Even if cooperators happen to form tight
clusters, they cannot survive once they face defectors.
Finally, the cooperators eventually extinguish
whatever values $p$ takes.
\end{description}

\begin{figure}
\begin{center}
\includegraphics[height=2in,width=2in]{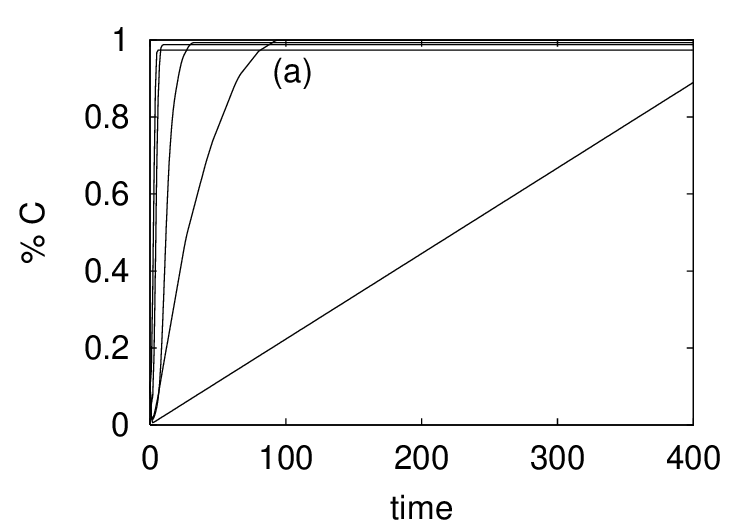}
\includegraphics[height=2in,width=2in]{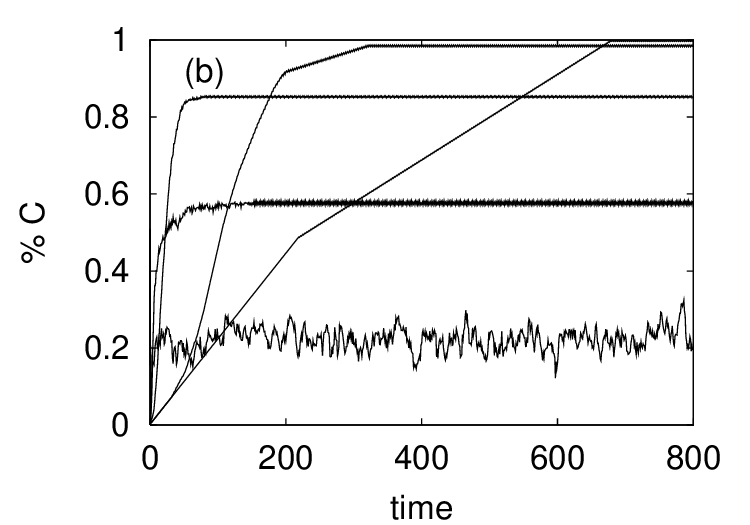}
\includegraphics[height=2in,width=2in]{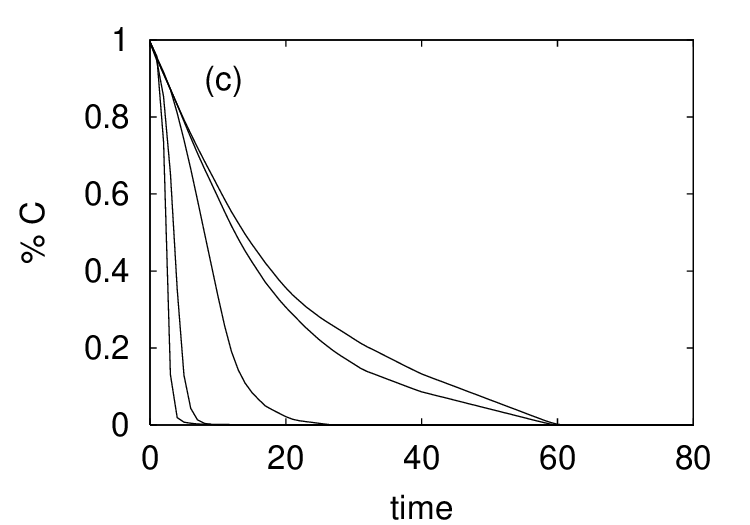}
\includegraphics[height=2in,width=2in]{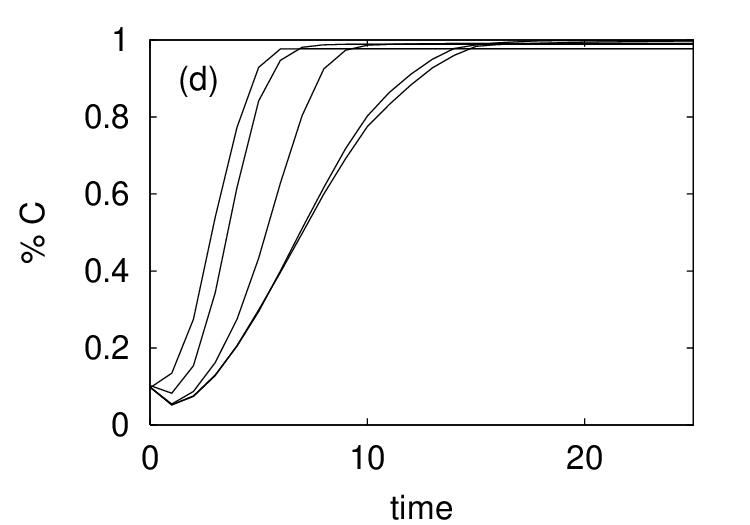}
\includegraphics[height=2in,width=2in]{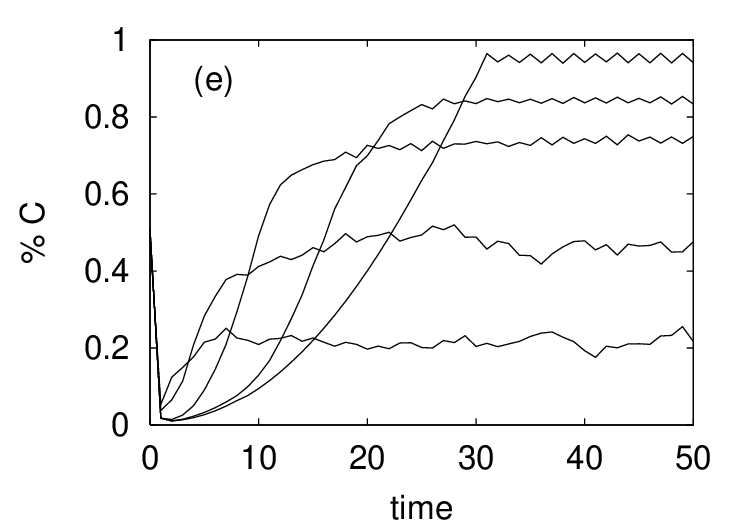}
\includegraphics[height=2in,width=2in]{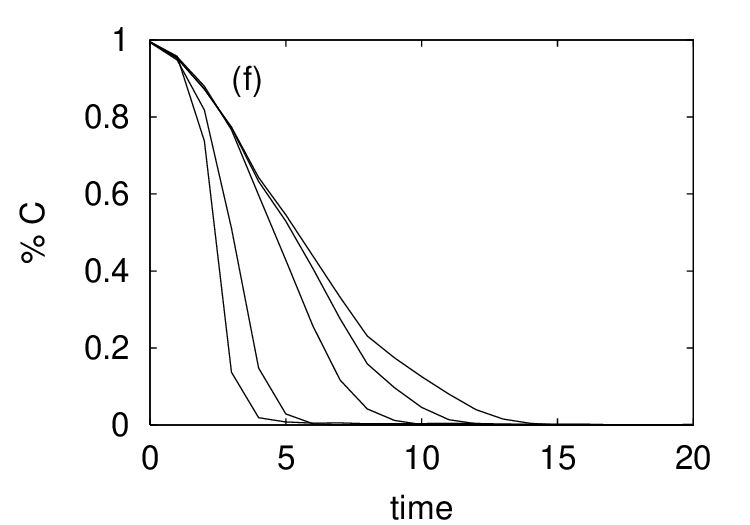}
\caption{The proportion of cooperators in the spatial PD with $d=1$
(a,b,c) and $d=2$ (d,e,f).  We set $(b,c(0)) = (1.1, 0.1)$,
$(b,c(0))=(1.7,0.5)$, $(b,c(0)) = (3.0, 0.995)$, for (a,d), (b,e), and
(c,f), respectively. The initial conditions are set randomly and
independently for all the players. The results for $p=0$, $p=0.001$,
$p=0.01$, $p=0.1$, and $p=0.8$ are shown, with more rapidly converging
lines corresponding to larger values of $p$.}
\label{fig:main}
\end{center}
\end{figure}

In regime (ii), which is of our particular interest, the
clustering coefficient and the number of cooperators are in close
relation.  Actually, the clustering property of networks is related to
the measure of assortment $r$ which, for example, represents the
relatedness coefficient associated with kin selection \cite{Boyd88}.
The proportion $r$ of the total players are supposed to side with the
decision of a reference player, and cooperators are more likely to
survive in more assortative populations. In accordance with this
observation, simple calculation leads to $p=1-r$.  Figures
\ref{fig:main}(b) and \ref{fig:main}(e) show temporal profiles of
the proportion of cooperators $c(n)$ after $n$ rounds in this regime
for $d=1$ and $d=2$, respectively. We set $b=1.7$ and the initial
condition $c(0)=0.50$.  We find more cooperators for smaller values of
$p$. The cooperators effectively form
clusters and occupy a large part of the network for small
$p$, while cooperators and defectors are balanced when $p$ is larger.
However, the convergence is faster for smaller $L(p)$, that is,
for larger $p$. 
Figures~\ref{fig:main}(b) and \ref{fig:main}(e) together with
further parameter search reveal that rapid
convergence to the states with as many cooperators as in the case of $p=0$,
which might be desirable from a 
social point of view, is not realized even in the
small-world schemes. This suggests that the range of $p$ yielding
$C(p)$ large enough for the maintenance of total governance of
cooperators and that guaranteeing the rapid convergence to the equilibrium
proportion of cooperators do not exactly overlap.
However, in this situation, we observe rapid convergence to the
proportion of cooperators just slightly less than that for $p=0$.
Small-world properties are
generally found in social networks
\cite{Albert,Kim,Wattsbook,Watts,Newman01}, and our results suggest that
small-world networks may be socially preferred.

Rapid convergence to equilibria is more clearly understood in regimes
(i) and (iii), with \FIGS\ref{fig:main}(a) and 
\ref{fig:main}(d) (or \ref{fig:main}(c) and \ref{fig:main}(f))
for regime (i) (or regime (iii)) showing the convergence to the
all-$C$ (or all-$D$) states. Comparing the results for these two
regimes, we cannot comment on whether small $L(p)$ is `good'
or `bad' in terms of payoff.
Small values of $L(p)$ just guarantee fast convergence to
final states whatever they are \cite{Kim}, and this fact agrees
with our daily experience. In these cases with extreme values of $b$,
the players are just imitating the neighbors in a coordinated manner,
and this game dynamics can be considered as a simple epidemic model
with permanent-removal dynamics that consists only of the susceptible
and infectious populations \cite{Wattsbook}. In this situation, the
infectious individuals, which is $C$ for regime (i) and $D$ for regime
(iii), devour the susceptible individuals, and this process is faster
for smaller $L(p)$. As a remark, convergence to chaotic states for
$1.8<b<2.0$
\cite{Nowak92,Nowak93ijbc} is also rapid with larger $p$ while the size of
$\lim_{n\to\infty} c(n)$ is severely sacrificed compared with the case
when $b=1.7$ (data not shown).

The comparison of \FIGS\ref{fig:main}(b) and \ref{fig:main}(e)
reveals that rapid convergence to the equilibria with 
considerable cooperators is more emphasized for $d=1$ than for $d=2$. This is
because the dynamic ranges of $C(p)$ and $L(p)$ with respect to $p$
are wider for $d=1$, which pronounces the differences in
the speed of convergence and the final proportions of cooperators.
Actually, if the number of neighbors is fixed, networks with less
dimensions have wider dynamical ranges of $C(p)$ and $L(p)$. Let us suppose
$k=8$. Then straightforward calculations show $C(0)=9/14$ for
$d=1$ and $C(0)=3/7$ for $d=2$, whereas $C(1)=k/n$ for both cases. Similarly,
$L(0)\cong n/2k$ for $d=1$, and
$L(0)\cong \sqrt{n}/2k$ for $d=2$, whereas $L(1)\cong \log(n)/\log(k)$
for both dimensions \cite{Watts}.
For better presentations of the results, we set $d=1$ in the remainder
of this letter. However, qualitatively similar results are obtained
for $d=2$ as well.

\begin{figure}
\begin{center}
\includegraphics[height=2in,width=2in]{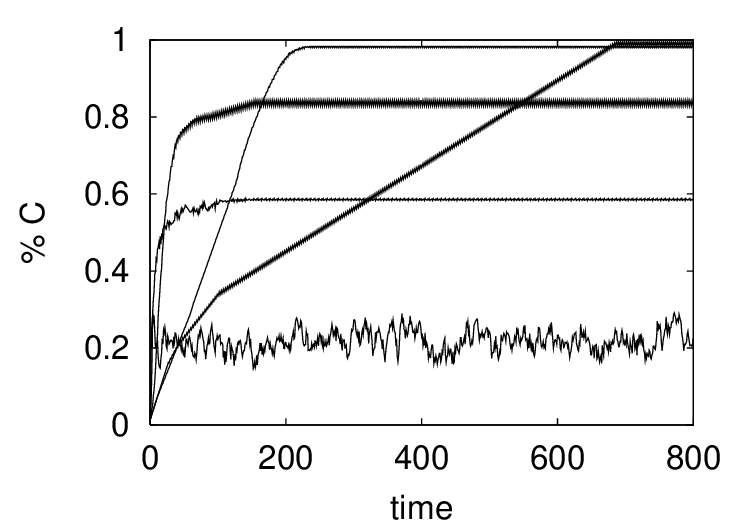}
\caption{The proportion of cooperators with clustered initial conditions.
We set $d=1$ and $(b,c(0)) = (1.7, 0.01)$.}
\label{fig:clus_init}
\end{center}
\end{figure}

To demonstrate generality of our results, let us note
the initial sharp drop of
$c(n)$ in \FIGS\ref{fig:main}(b) and \ref{fig:main}(e)
due to the random initial conditions. In the beginning,
defectors devour cooperators that do not form clusters. Then only
tight clusters of cooperators survive and then begin to grow.
A more relevant initial condition is clusters of cooperators in the
sea of defectors. Therefore we examine the dynamics starting with randomly
placed clusters each of which contains
nine successively aligned cooperators in the background of defectors. The
positions of the clusters are determined randomly and independently.
The results for $b=1.7$ and $c(0)=0.01$ shown in 
\FIG\ref{fig:clus_init} are similar to those in
\FIG\ref{fig:main}. The same is also true for regime
(i) and for $d=2$ (data not shown).

\begin{figure}
\begin{center}
\includegraphics[height=2in,width=2in]{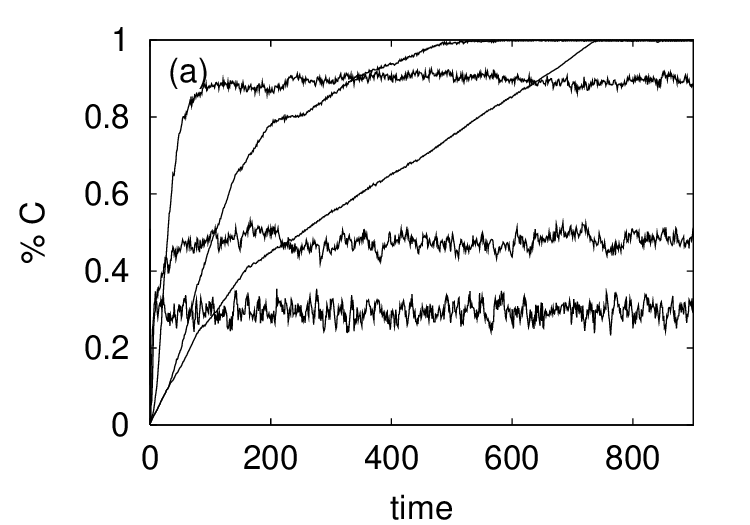}
\includegraphics[height=2in,width=2in]{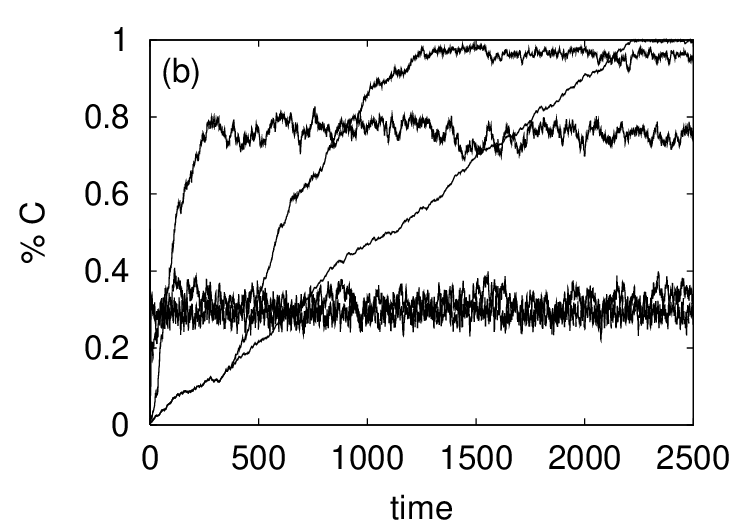}
\caption{The proportion of cooperators in the case of $d=1$
with independent initial conditions and noise.
We set $(b,c(0),m) = (1.6, 0.5, 20)$ and $(b,c(0),m) = (1.55, 0.5,
10)$ for (a) and (b), respectively.}
\label{fig:noisy}
\end{center}
\end{figure}

Finally, we consider effects of noise. In the real world, decision of
each player can accompany errors because information
available for decision making is imperfect 
and the behavior can be inherently
unpredictable to some extent.
Now we introduce stochastic decision rules in which
the probability that a player chooses $C$ in the next round
is $\sum^k_{i=0} A^m_i s_i \bigg/
\sum^k_{i=0} A^m_i$ \cite{Nowak94pnas}.  Here $A_i$ is the payoff of
the $i$th neighbor, with $i=0$ corresponding to the player itself.
When the current strategy of the $i$th neighbor is
$C$ (or $D$), $s_i$ takes one (or zero).
The value of $m$ specifies the degree of
stochasticity. The deterministic case corresponds to $m=\infty$, and
smaller $m$ indicates more randomness.
The results for $m=20$, $b=1.6$, $c(0)=0.5$ and those for
$m=10$, $b=1.55$, $c(0)=0.5$ are shown in \FIGS\ref{fig:noisy}(a)
and \ref{fig:noisy}(b), respectively. Similarity of the results 
to the deterministic cases in spite of apparently large amount of noise
implies the robustness of the results
against noise. However, larger noise requires smaller $b$
for the same level of cooperation to
be maintained because noise spoils the effect of clustering
\cite{Nowak94pnas}.
Noise also lengthen the transient time
as manifested in \FIG\ref{fig:noisy}(b).
For still larger amount of noise with smaller $m$, we do not
obtain quasi-convergent behavior \cite{Nowak94pnas}.

\section{Conclusions}

We have shown that, in the spatial PD, three different kinds of dynamics and
dependence on the network structure exist with respect to $b$.  For
intermediate values of $b$, small-world architecture realizes a
quasi-optimal behavior in the sense of rapid convergence to a good
equilibrium.  Here the `goodness' is implicitly measured by the
following hierarchy of states. The best consequence is fast
convergence to a equilibrium with many cooperators, the second best is
slow convergence to this type of equilibria. The worst is fast
convergence to a population with many defectors, and the second worst
is slow convergence to such a state.

We have also investigated the effects of dimensionality, clustered
initial conditions, and dynamical noise. The results are qualitatively
the same for these cases, but networks with smaller dimensions generally
lead to more enhanced effects of the small-world property because
of the wider dynamic ranges of $C(p)$ and $L(p)$.  In nongeometric
social networks such as ecological networks, food webs, and friendship
networks, the dimension of the underlying substratum is not really
known whichever value of $p$ is appropriate. This is in contrast with
networks based on physical locations such as neural networks and power
grids. Further investigation of the dimensionality effect is our
future problem.

\section*{Acknowledgments}

This work is supported by the Japan Society for the Promotion of Science, by the Advanced and Innovational Research program in Life Sciences from the Ministry of Education, Culture, Sports, Science, and Technology, the Japanese Government, and by Superrobust Computation Project in 21st Century COE Program on Information Science and Technology Strategic Core.

\end{document}